\documentclass[twocolumn,showpacs,preprintnumbers]{revtex4}
\usepackage{graphicx}
\usepackage{dcolumn}
\usepackage{bm}
\usepackage{multirow}
\usepackage{amsfonts}
\usepackage{amsmath}
\usepackage{amssymb}
\usepackage{revsymb}
\usepackage{longtable}

\begin{document}

\title{Manipulation of Dark States and Control of Coherent Processes with
Spectrally Broad Light}
\author{M.~Auzinsh$^{1,2}$}
\email{Marcis.Auzins@lu.lv}
\author{N.~N.~Bezuglov$^2$}
\thanks{Permanent address: physical faculty, Fock Institute of Physics,
St.Petersburg State University, 198904 St. Petersburg, RUSSIA}
\author{K.~Miculis$^2$}
\affiliation{$^1$University of Latvia, Dept. of Physics and Mathematics, 19 Rainis
Boulevard, LV-1586 Riga, Latvia}
\affiliation{$^2$Laser Centre, University of Latvia, LV-1586 Riga, Latvia}
\date{Version 4 \today }

\begin{abstract}
The formation of dark states under interaction of degenerate atomic states
with incoherent broadband radiation (white light) is discussed. A simple coupling
scheme in a three level $\Lambda $-system, which allows
both qualitative and quantitative analysis is discussed. We found a stationary solution
of the optical Bloch equations in a broad excitation line approximation
that describes the dynamics of the atom--white light interaction and demonstrated
its identity to a conventional dark state created with coherent laser
fields. We than examine the efficiency of the population transfer induced
by broadband radiation in a model $\Lambda $-system and revealed that high efficiency
(attaining 100\%) of stimulated Raman adiabatic passage-like processes can be achieved
with certain temporal control of light polarization. The
corresponding criterion of adiabaticity was formulated and justified by means of
numerical simulations.
\end{abstract}

\maketitle




\section{\label{intro} Introduction}

Coherent laser fields can be used to manipulate atomic and molecular quantum
states in order to create coherent superpositions of quantum states, which are
of interest in many important fields of research including laser cooling \cite{Nobel},
ultra cold matter \cite{Metcalf}, dark states \cite{Enio},
electromagnetically induced transparency \cite{Fle05}, and laser driven states
\cite{Klaas}. These techniques have many important applications at the
forefront of technology and industry, including in such areas as lithography
\cite{Wal01}, quantum information \cite{Bou2000}, quantum chemistry \cite%
{Sha03}, and others. Typically, control of quantum states is implemented
within $\Lambda $-type and $V$-type systems driven by two (or more) laser
fields. In the case of $\Lambda $-type excitation, dark (or population trapping)
states ($D$-states) can be formed, which become transparent for photons and
thus cease interacting with light \cite{Enio}. In contrast, another
type of superposition of coherent states, bright states ($B$-states), becomes
more absorbing under the laser light action \cite{Aln01,Ren01,Pap02}. The
dark states are useful in laser cooling processes \cite{CoTa, Metcalf},
optical pumping \cite{Happer, Marcis, We}, \textquotedblleft lasing without
inversion\textquotedblright\ \cite{Koc88,Scu89}, and others. One particular
application of dark states is in the method of Stimulated Adiabatic Raman
Passage (STIRAP), which can be used to couple an initial and final state to
a common intermediate state and transfer atomic populations between initial
and final state without loss \cite{Klaas}. This method has traditionally
required highly stabilized laser fields that are strongly phase correlated.
But are these requirements really so strict as traditionally
thought? In this article we discuss a scheme which allows one to achieve
high degree of control of quantum states by means of broad-band (i.e.,
basically white) light from a single light source.

The basic equation that describes coherent processes in light--atom
interaction under the density matrix formalism is the Liouville equation, which in
atomic physics often is referred to by the term \textquotedblleft optical Bloch
equations\textquotedblright. When these equations describe a system
illuminated by spectrally broad light, they formally allow a steady-state
solution that has the structure of a pure quantum state similar to the
solution for dark states in monochromatic laser fields. It turns out that
spectrally broad light can form dark states thanks to a beneficial
cancellation of photons of different frequencies. This result evidently
opens new possibilities for controlling atomic states with incoherent light
sources. As a specific case, we will discuss in detail the problem of
transferring a population of atoms with spectrally broad light from an
initial, stable discrete state to a desired target state, without loss of
population.

The paper is organized as follows. In Section \ref{problem} we shall review
the most optimal scenario for population transfer; we discuss the nature of
coherent dark states under monochromatic excitation and outline the main
factors of the atom--coherent light interaction that can destroy the $D$%
-states by mixing them with bright states. In Section \ref{phase} we shall
consider cases that limit the control over coherence, such as two lasers
that are only partially coherent. We will then be in a position to formulate
the arguments for and against the possibility of using spectrally broad
light to manipulate quantum states. The answer will be obtained in Section %
\ref{dark}. In Section \ref{equation} we shall analyze the master equation
for the density matrix that describes the dynamics of the atom--spectrally
broad light interaction \cite{CT, Blushs}. In Section \ref{states} we shall
demonstrate the existence of $D$-states under spectrally broad light
excitation and coherent processes with broad band excitation. Then, in
Section \ref{whiteSTIRAP} we shall present the results of numerical
simulations that demonstrate that population transfer by spectrally broad
light can achieve an efficiency of 100 $\%$. The effects of detuning from
the two-photon resonance will be analyzed in Section \ref{Zeeman}, and
numerical simulations of the corresponding two photon line shapes will be
presented. In Section \ref{concl} we will summarize the results and discuss
the possibilities of coherent processes in atoms in the case of spectrally broad
light excitation. Finally, in order to justify some statements related to $D$%
-states and STIRAP in the case of spectrally broad light sources, an analogy
between coherent and incoherent process for quantum state control is
established in the Appendix.

\section{\label{problem} Formulation of the problem: adiabatic passage
within bound states}

\begin{figure}[tbp]
\begin{center}
\label{Fig1} \includegraphics[width=5cm]{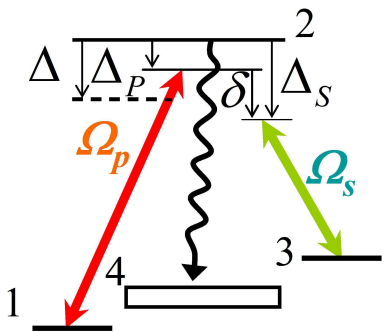}
\end{center}
\caption{Schematic diagram of energy levels, laser Rabi frequencies, and
detunings. The initial population is embedded on level 1.}
\end{figure}

We shall concentrate ourselves on the problem of how to transfer populations
of atoms from an initial level $1$ to some target level $3$ (see Fig.~1)
without loss of population. We first briefly recall the case of two
coherent, \textit{monochromatic} pulsed lasers, a Pump laser and a Stokes
laser, with fixed frequencies $\omega _{P}$, $\omega _{S}$ and corresponding
Rabi frequencies $\Omega _{P}(t)$, $\Omega _{S}(t)$ \cite{Klaas}. It is
clear that, in any atom manipulation scheme, one should avoid involving the
unstable upper-lying state $2$, because from this state population could
flow into other unwanted states, which are schematically depicted in
Figure~1 as a single level $4$. Our first task is to find how to create a
wave functions $\Psi _{D}=C_{1}\Psi _{1}+C_{3}\Psi _{3}$ as a linear
combination of the two low-lying states $1$ and $3$, which is not coupled to
the excited state $2$. In other words, if the Hamiltonian is $H=H_{0}+V$,
where $V$ describes the coupling of the system levels with light, we require
that the matrix element $\langle \Psi _{2}|V|\Psi _{D}\rangle $ is zero.


\subsection{\label{dark_bright} Dark and bright states}


In the rotating wave approximation (RWA), the total Hamiltonian $H = H_0 + V$
of the system depicted in Fig.~1 has the well known form \cite{Leonid}:

\begin{equation}  \label{eq1}
H = \hbar/2 \left[
\begin{array}{ccc}
2\Delta_P & \Omega_P(t) & 0 \\
\Omega_P(t) & 0 & \Omega_S(t) \\
0 & \Omega_S(t) & -2\Delta_S \\
&  &
\end{array}
\right]
\end{equation}

\noindent in the basis of the bare states $\Psi _{i}$ $(i=1,2,3)$. The
Hamiltonian $H_{0}$ corresponds to a free atom and determines the bare state
energies $\varepsilon _{i}$. We choose as the zero level of energy the value
$\varepsilon _{2}$ for the excited state $2$. The quantities $\Delta _{P,S}$
in the diagonal elements give the laser detunings ($\Delta _{P}=\omega
_{P}-\omega _{21}$; $\Delta _{S}=\omega _{S}-\omega _{23}$) from the Bohr
frequencies $\omega _{21}$, $\omega _{23}$ of the corresponding optical
transitions (see Fig.~1). The Rabi frequencies $\Omega _{P}(t)$, $\Omega
_{S}(t)$ are determined from the coupling term $V$ of the Hamiltonian. We
neglect to mention here any relaxation terms and leave their proper
discussion to Sec.\ref{dark}.

The required solution $\Psi_D(t)$ of the equation $\langle\Psi_2|V|\Psi_D%
\rangle = 0$ reads \cite{Klaas}:

\begin{equation}
\begin{array}{l}
\Psi_D = \cos\Theta(t)\Psi_1 - \sin\Theta(t)\Psi_3 \; ; \\
\sin\Theta = \Omega_P/\Omega_{eff} \; ; \quad \Omega_{eff} =\sqrt{%
|\Omega_P|^{2}+|\Omega_S|^{2}}%
\end{array}
\label{eq2}
\end{equation}

\noindent and is known as a dark state. Since $\Psi_D(t)$ does not share the
population with the excited state, it does not radiate directly itself. Note
that the mixing angle $\Theta$ gives a convenient measure of population
sharing between stable states: the value $\Theta=0$ corresponds to a
population that resides entirely in state $1$, while $\Theta=\pi/2$
describes a population that has been transferred entirely to the target
state $3$.

\begin{figure}[tbp]
\begin{center}
\label{Fig2} \includegraphics[width=7cm]{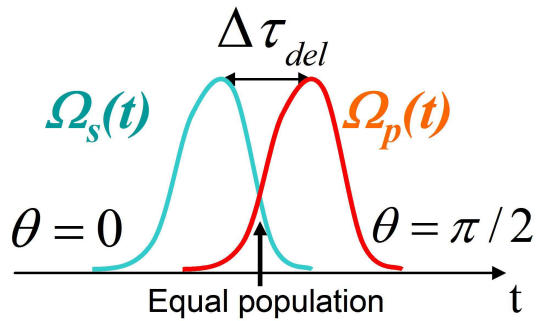}
\end{center}
\caption{A sequence of laser pulses. The arrow indicates a moment when the
populations of levels 1 and 3 are equal.}
\end{figure}

Since controlling the population means controlling the mixing angle, as can
be seen from Eq.~(\ref{eq2}), an efficient transfer of population can be
achieved by organizing a sequence of dark states with the mixing angle
varying from $\Theta =0$ (the initial state $1$ populated) to $\Theta =\pi /2
$ (the target state $3$ populated). In this sequence of states, the state
vector $\Psi (t)$ should be maintained as the current dark states $\Psi
_{D}(t)$ all the time. An experimental implementation of such a scenario is
depicted in Fig.~2. The pulse sequence that provides the desired rotation of
the mixing angle seems counter-intuitive \cite{Sho95}: the pump laser pulse
arrives after the Stokes laser pulse! What happens is that the Stokes laser
prepares (dresses) the transition $2\rightarrow 3$ for accepting the
population, which is delivered by the pump laser \cite{Klaas}. It is noteworthy that
the states $1,3$ share the population equally at the moment when both
lasers' Rabi frequencies are equal.

We now examine different unwanted factors that can restrict the efficiency
of the desired population transfer. For this purpose, we consider another
convenient concept, the so called bright state ($B$-state) $\Psi_B$ \cite%
{Klaas}:

\begin{equation}
\begin{array}{l}
\Psi_B = \sin\Theta(t)\Psi_1 + \cos\Theta(t)\Psi_3 \; ; \\
\langle\Psi_2|V|\Psi_B\rangle/\hbar = \Omega_{eff}%
\end{array}
\label{eq3}
\end{equation}

\noindent which is orthogonal to the dark state $\Psi _{D}$. Although the $B$%
-state also does not contain the excited state $2$, it is always mixed with
the excited state by the light fields, as can seen from the corresponding
matrix element presented in Eq.~(\ref{eq3}). Having been coupled to the
excited state, the $B$-state thus is power broadened, and this
coupling leads to radiation from state $2$, which results in unwanted population
losses to the marginal levels $4$ due to spontaneous transitions (see
Fig.~1).

As was mentioned above, although the $D$-state is not coupled to state $2$
and does not lead to radiation from that state, nevertheless it may be
coupled with the bright state. The mixing frequency between the bright and
dark states during the system's temporal evolution reads \cite{Leonid}:

\begin{equation}
\langle \Psi _{B}|\frac{H}{\hbar }+i\frac{\partial }{\partial t}|\Psi
_{D}\rangle \equiv i\frac{d\Theta }{dt}-\frac{\delta }{2}\sin 2\Theta
\label{eq4}
\end{equation}%
where $H$ is the Hamiltonian (1), and the temporal derivative corresponds to
non-adiabatic linkage between states. Two important requirements follow from
relation (\ref{eq4}). First, to preserve the dark state, changes of the
mixing angle (see Eq.~\ref{eq2}) should be slow enough, or adiabatically
organized. The corresponding criterion is given via the inequality $d\Theta
/dt\ll \Omega _{eff}(t)$, which yields, after integration over $t$ \cite%
{Klaas}:

\begin{equation}  \label{eq5}
\int\limits_{ - \infty} ^{\infty} dt\Omega_{eff}(t) \gg \Delta\Theta = \pi/2
\; .
\end{equation}

\noindent It is seen that the applied laser pulses should be stronger than
the $\pi$-pulses. The second important requirement concerning the mixing of $%
B$- and $D$-states is that the difference $\delta = \Delta_S - \Delta_P$
between laser detunings should be small. This difference $\delta$ (see
Fig.~1) is often called the double-photon detuning, and it opens a pathway
for unwanted population flow, which may dramatically destroy the $D$-state.
A detailed study of the efficiency of STIRAP-like processes as a function of
$\delta$ (the so called two-photon line shape) may be found in \cite%
{Romanenko}. Note that the one photon detuning $\Delta$, determined as $%
\Delta = 1/2(\Delta_S + \Delta_P)$ (see Fig.~1), does not enter itself
into the mixing matrix element in Eq.~(\ref{eq4}), which explains the weak
influence of $\Delta$ on the population transfer \cite{Klaas, Romanenko}.


\subsection{\label{phase} Phase-diffusion effects for partially coherent
fields}


Up to now we dealt with coherent radiation. However, in the real world,
lasers typically are subject to vibrations and other environmental
influences that cause phase diffusion and result in only partial coherence
of the laser fields. Phase diffusion effects were studied in detail by \cite%
{Leonid} using the phase diffusion model, according to which the random walk
of laser frequencies varies chaotically both the double-photon detuning $%
\delta$ and the single-photon detuning $\Delta$. Drift of the one photon
detuning is not detrimental to STIRAP, but strong $\delta$-chaotic jumps
dramatically decrease the STIRAP efficiency. However, the authors of paper \cite%
{Knight} pointed out an important exception: if the radiation in both laser
fields has the same source, a beneficial cancelation of the phase
fluctuations may occur. Since laser phases are varying equally, the value of
$\delta$ remains equal to zero.

\section{\label{dark} Dark states in spectrally broad light}

With the above preliminaries, we are now ready to determine if quantum
states can be controlled by means of spectrally broad light instead of
coherent lasers. From Section \ref{phase} it is clear that in case of
spectrally broad light one has to use a single light source. Otherwise, if
two distinct uncorrelated sources of broad-band light were to be used, every
dark state would be depopulated by mutual, multiple incoherences among the
sources. In our analysis we deal with a fluctuating electric field $\mathbf{E%
}(t)$ that has a well defined elliptical polarization:

\begin{equation}
\begin{array}{l}
\mathbf{E}(t)={\rm Re}\mathbf{E}_{0}(t)\exp (-i\omega _{0}t)\varepsilon
(t)\;; \\
\langle \varepsilon (t_{1})\varepsilon ^{\ast }(t_{2})\rangle
=A(t_{1}-t_{2})\;,%
\end{array}
\label{eq6}
\end{equation}

\noindent where $\omega _0$ is the carrier frequency of the light. We assume
that the fluctuating part $\varepsilon(t)$ of the light is a scalar,
dimensionless, random, complex function of unit modulus $|\varepsilon|=1$ with
the broadband correlation function $A(t_1 - t_2)$. As a result, the spectral
distribution $P(\omega)$ of the light \cite{Feynman}

\begin{equation}
P(\omega )=\frac{1}{\pi }Re\int_{0}^{\infty }dt\exp (-i\omega t)A(t)
\label{eq7}
\end{equation}%
is a smoothly varying function within the spectral interval $\Delta \omega $
of interest. The application of adiabatic elimination procedure
implies the following requirement $\tau \Delta \omega \gg 1$ for the
characteristic duration $\tau $ of the matter/light interaction, which is directly
related to the time of switching on and switching off the light beam
\cite{Stenholm,Blushs}. When
the interaction takes place during a very short time interval, i.e., when $%
\tau $ becomes very small, the correlation function $A(t_{1}-t_{2})\sim
\delta (t_{1}-t_{2})$ should correspond to spectrally broad light (here $%
\delta $ is Dirac delta-function). In the case of slow processes, i.e.,
in the case of adiabatic control of quantum states, the spectral
interval $\Delta \omega $ may be of finite size. It is important to note
that the frequency $\omega $ in Eq.~(\ref{eq7}) denotes a measure of the
frequency shift from the center $\omega _{0}$ of the light spectra.

\begin{figure}[tbp]
\begin{center}
\label{Fig3} \includegraphics[width=5cm]{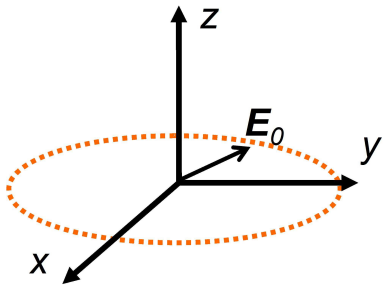}
\end{center}
\caption{Light polarization.}
\end{figure}

In contrast, the envelope vector function $\mathbf{E}_{0}(t)$ has regular
behavior and determines the light polarization. If one chooses the direction
of light propagation as the $\emph{z}$-axis (see Fig.~3), the
polarization ellipse lies in the $(\emph{x,y})$-plane. It is convenient to
work with the polarization elements $E^{(\pm 1)}$ represented by spherical
components $\mathbf{e}_{\pm }$ \cite{Landau} of vector $\mathbf{E}_{0}$:

\begin{equation}
\begin{array}{l}
\mathbf{E}_{0}(t)=E^{+1}(t)\mathbf{e}_{+}+E^{-1}(t)\mathbf{e}_{-}; \\
\mathbf{e}_{\pm }=\mp (\mathbf{e}_{x}\pm i\mathbf{e}_{y})/\sqrt{2}\;.%
\end{array}
\label{eq8}
\end{equation}%
When $E^{(\pm 1)}$ are real, the polarization of light is such that the main
semi-axes of the ellipse are oriented along the $\mathbf{e}_{x}$ and $%
\mathbf{e}_{y}$ axes. If $E^{(\pm 1)}$ are complex numbers, the difference
between their phases determines the double rotation angle of the ellipse in
the $\emph{(x,y)}$-plane.

The quantum states may be controlled through an appropriately chosen,
time-dependent variation of the light polarization. As an example we
consider the simple interaction scheme presented in Fig.~4: a single
broad-band light beam excites a two-level system. The excited state ($e$%
-state) has the angular momentum $l=0$ and consists of one Zeeman component $%
m^{\prime }=0$. The ground state ($g$-state) posses angular momentum $l=1$
and therefore has three components, one of which $(m=0)$ is not involved
in the interaction because of the chosen light polarization plane. The
light's central frequency $\omega _{0}$ is assumed to be in resonance with
the $g\rightarrow e$ transition. The quantization axis is oriented along the
\emph{z}-direction. As is apparent, the spherical component of light (\ref%
{eq8}) couples independently the transitions $m=\pm 1\rightarrow m^{\prime
}=0$. In fact, the magnetic sublevels involved in this interaction
effectively form a three-state $\Lambda $-scheme.

In our model, we assume that at $t=-\infty $ only the $m=-1$ component is
populated. Our aim is to analyze the efficiency of the STIRAP-like process
that could transfer the population from $m=-1$ to $m=+1$. This population
transfer can be accomplished by applying a sequence of light pulses with
Rabi frequencies $\Omega _{\mp }$ (see Fig.~2) by varying (by changing light
polarization) the relative strength of the polarization components $E^{(\pm
1)}$:

\begin{equation}
\begin{array}{l}
E^{(\pm 1)}(t)=E_{S}\exp (-(t-\Delta \tau _{\mp })^{2}/2\tau ^{2})\;; \\
\Omega _{\mp }(t)=E^{(\pm 1)}(t)\langle m^{\prime }=0|\mathbf{e}_{\pm }\cdot
\mathbf{d}|m=\mp 1\rangle /\hbar \;.%
\end{array}
\label{eq9}
\end{equation}%
Here $\mathbf{d}$ denotes the atomic dipole moment. The Rabi frequencies $%
\Omega _{\pm }$ correspond to the strengths of the coupling interaction
between the levels $m=\pm 1$ and $m^{\prime }=0$, which is induced by the
light's spherical components $E^{\mp }$ (see Fig.~4). The parameter $\tau $ in
the arguments of the exponential factors determines the interaction time,
while the parameters $\Delta \tau _{\pm }$ give the temporal shifts of the
applied impulses. It is worth emphasizing an important feature of the
scheme presented here. Clearly, each frequency $\omega $ of the light beam
stimulates both transitions $m=\pm 1\rightarrow m^{\prime }=0$ with
effective partial Rabi frequencies $\bar{\Omega}_{\pm }(\omega )=\Omega
_{\pm }\sqrt{P(\omega )}$ and results in the appearance of dark states (\ref%
{eq2}). Because the corresponding mixing angle $\sin \Theta =\bar{\Omega}%
_{+}/\sqrt{\bar{\Omega}_{+}^{2}+\bar{\Omega}_{-}^{2}}$ turns out to be
independent of $\omega $, the photons prepare a unique dark state, which therefore
is not coupled to the upper excited state $m^{\prime }=0$.

Here, however, new additional unwanted factors in the formation of the dark
state arise. Indeed, spectrally broad light actually consists of many
uncorrelated photons with different frequencies $\omega _{0}+\omega $. A
photon with the fixed frequency $\omega _{0}+\omega $ that excites the
transition $m=+1\rightarrow m^{\prime }=0$ (the Rabi frequency $\bar{\Omega}%
_{+}(\omega )$) combines with a variety of photons of frequencies $\omega
_{0}+\widetilde{\omega }$ that excite the transition $m=-1\rightarrow
m^{\prime }=0$ (the Rabi frequency $\bar{\Omega}_{-}(\widetilde{\omega })$).
We could expect the presence of many nonzero two-photon resonance detunings $%
\delta =\omega -\widetilde{\omega }$ to lead to strong mixing between the
dark and bright states (see Eq.~(\ref{eq4})), i.e. to a fast destruction of
the dark state. However, there is one favorable circumstance: it is possible
to distribute $\tilde{\omega}$-frequencies into pairs $\tilde{\omega}_{1,2}$
with opposite two-photon detuning values $\delta _{1}=-\delta _{2}$, so that
the average value $\langle \delta _{1,2}\rangle $ over the pair becomes
zero. As a result, it may be possible to compensate the pair contribution in
the mixing between dark and bright states. The total rate of unwanted
population loss due to coupling to the $B$-state could be still reduced
to zero despite the presence of many light frequencies. To verify this hypothesis,
we need a robust treatment of the dynamics of the system.

\begin{figure}[tbp]
\begin{center}
\label{Fig4} \includegraphics[width=5cm]{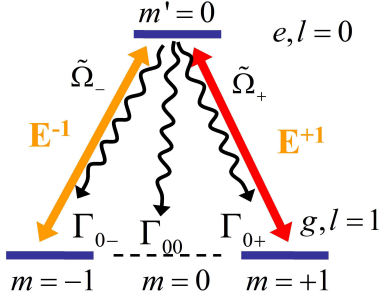}
\end{center}
\caption{Energy level diagram}
\end{figure}


\subsection{\label{equation} Basic equation for the density matrix under
coupling with spectrally broad light}


The evolution of the system should be studied within the framework of the
density matrix $\rho _{ij}$ formalism. At first glance, the problem seems to
be intractably complicated because of the presence of multi-chromatic light.
Fortunately, the broad spectrum of applied spectrally broad light results in
an adiabatic elimination of the optical coherences \cite{Stenholm, Blushs}
in the optical Bloch equations, i.e., $\rho _{eg}$ in our case. Briefly, under
the action of spectrally broad light, the effective lifetime $\tau _{ex}\sim
1/\Delta \omega $ of induced optical dipoles described by $\rho _{eg}$
appears to be very short in comparison with the interaction time $\tau $.
The optical dipoles adiabatically follow only the $\rho _{ee}$, $\rho _{gg}$
elements (Zeeman coherences). The diagram techniques for solving the evolution of
the density matrix developed by Konstantinov and Perel \cite{Perel} and
later by Keldish \cite{Keldish} justify the results that were first
empirically obtained by Claude Cohen-Tannoudji \cite{CT} in the form of rate
equations for the Zeeman coherences $\rho _{ii}$ under the broad-line
approximation. A detailed study of various problems under broad-line
approximation may be found in \cite{Blushs}. In particular, the equations
presented in \cite{Blushs} and adopted for our system (Fig.~4) may be easily
reduced to the following system of equations, which describes the
populations of $e$-state ($\rho _{00}^{^{\prime }}$), $g$-state ($\rho _{++}$%
, $\rho _{--}$) and the off-diagonal ($\rho _{+-}$, $\rho _{-+}$) elements
between $g$-state components $m=\pm 1$ (Zeeman coherences):

\begin{eqnarray}
\frac{d}{dt}\rho _{00}^{^{\prime }}=-\left( |\widetilde{\Omega }_{+}|^{2}+|%
\widetilde{\Omega }_{-}|^{2}+\Gamma _{0}\right) \rho _{00}^{^{\prime }}+|%
\widetilde{\Omega }_{+}|^{2}\rho _{++}\;+ &&  \notag \\
|\widetilde{\Omega }_{-}|^{2}\rho _{--}+2Re\widetilde{\Omega }_{-}\widetilde{%
\Omega }_{+}^{\ast }\rho _{-+}\;;\qquad \qquad \qquad  &&  \label{10a} \\
\frac{d}{dt}\rho _{++}=-|\widetilde{\Omega }_{+}|^{2}\rho _{++}+\left( |%
\widetilde{\Omega }_{+}|^{2}+\Gamma _{0+}\right) \rho _{00}^{^{\prime
}}\;-\qquad \quad  &&  \notag \\
Re\widetilde{\Omega }_{-}\widetilde{\Omega }_{+}^{\ast }\rho _{-+}\;;\qquad
\qquad \qquad \qquad \qquad \qquad  &&  \label{10b} \\
\frac{d}{dt}\rho _{--}=-|\widetilde{\Omega }_{-}|^{2}\rho _{--}+\left( |%
\widetilde{\Omega }_{-}|^{2}+\Gamma _{0-}\right) \rho _{00}^{^{\prime
}}\;-\qquad \quad  &&  \notag \\
Re\widetilde{\Omega }_{-}\widetilde{\Omega }_{+}^{\ast }\rho _{-+}\;;\qquad
\qquad \qquad \qquad \qquad \qquad  &&  \label{10c} \\
\frac{d}{dr}\rho _{-+}=\left[ 2i\omega _{L}^{{}}-\frac{1}{2}\left( |%
\widetilde{\Omega }_{+}|^{2}+|\widetilde{\Omega }_{-}|^{2}\right) \ \right]
\rho _{-+}\;-\qquad  &&  \notag \\
\frac{1}{2}\widetilde{\Omega }_{-}^{\ast }\widetilde{\Omega }_{+}\left( \rho
_{++}+\rho _{--}-2\rho _{00}^{^{\prime }}\right) \;.\quad \qquad \qquad  &&
\label{10d}
\end{eqnarray}%
Note that $\rho _{+-}=\rho _{-+}^{\ast }$.

The structure of system (\ref{10a})-(\ref{10d}) contains two types of terms.
First, the majority of its terms contain the populations and, hence,
represent a simple balance between population flow into different levels.
Second, the contribution of coherence effects is represented by the Zeeman
coherence element $\rho _{+-}$. Briefly, the spontaneous radiative constant $%
\Gamma _{0}$ gives the total rate of the excited state $m^{\prime }=0$ decay
into the low-lying levels. The values $\Gamma _{0i}$ ($i=-,0,+$) yield the
partial spontaneous transition rates from the $e$-state to the individual $i$%
-components of the $g$-state; clearly $\Gamma _{0}=\Gamma _{0-}+\Gamma
_{00}+\Gamma _{0+}$. In our particular case (Fig.~4), one has $\Gamma
_{0i}=\Gamma _{0}/3$ \cite{Sobelman}. The light polarization components $%
E^{\pm 1}$ generate transitions, including stimulated transitions, between
the excited and ground states. The efficiency of the coupling of states is
determined by the effective "frequencies" $\widetilde{\Omega }_{\pm }$,
which turn out to be expressed via the above introduced Rabi frequencies $%
\Omega _{\pm }(t)$ (\ref{eq9}):

\begin{equation}
\widetilde{\Omega }_{\pm }(t)=\hbar \sqrt{P(\omega \!=\!0)}\Omega _{\pm }(t)=%
\sqrt{P(\omega \!=\!0)}E^{\pm 1}(t)\Vert d\Vert /\sqrt{3}\;.  \label{11a}
\end{equation}

\noindent The parameter $\Vert d\Vert $ corresponds to the reduced dipole
matrix element \cite{Sobelman}, while the coefficient $\sqrt{3}$
arises from 3-$j$ symbol related to $m=\pm
1\rightarrow m^{\prime }=0$ transitions according
to the Wigner-Eckart theorem. Note that because of the factor $%
\sqrt{P}$, the dimension of $\widetilde{\Omega }_{\pm }$ is [$s^{-1/2}$].
The purely imaginary term in Eq.~(\ref{10d}) appears in the presence of an
external magnetic field that results in the Larmor
energy shift $m\hbar \omega _{L}$ of Zeeman components $m$, i.e., in  energy
splitting $2\hbar \omega _{L}$ between $m=-1$ and $m=+1$ components.

It is convenient to measure the scale of the coupling in units of $E_S$ (\ref%
{eq9}), so that the pulse sequences~(\ref{eq9}) acquire the form

\begin{equation}
\begin{array}{l}
\widetilde{\Omega }_{\pm }(t)=\widetilde{\Omega }_{0}\exp (-(t-\Delta \tau
_{\pm })^{2}/2\tau ^{2})\;; \\
\hbar \widetilde{\Omega }_{0}=\sqrt{P(\omega =0)}E_{S}\Vert d\Vert /\sqrt{3}%
\end{array}
\label{11b}
\end{equation}%
with $\widetilde{\Omega }_{0}$ again measured in $[\mathrm{s}^{-1/2}]$.

Coherent processes, as was mentioned above, are described by the dynamics of
the off-diagonal elements $\rho _{+-}=\rho _{-+}^{\ast }$. Because of these
off-diagonal elements the balance equations result in some important specific features,
such as, for instance, the existence of dark states. In particular:

(i) The system of Eqs.~(\ref{10a})--(\ref{10d}):
describes an open system because of the presence of a spontaneous
cascade into the uncoupled $m^{\prime }=0$ component of the $g$-state. In
accordance with Eqs.~(\ref{10a})--(\ref{10c})

\begin{equation}  \label{eq12}
\frac{d}{dt}(\rho^{^{\prime
}}_{00}+\rho_{++}+\rho_{--})=-\Gamma_{00}\rho^{^{\prime }}_{00}
\end{equation}

\noindent population flow into the uncoupled level depletes the population
of the coupled system. Correspondingly, the population $\rho^{}_{00}$ of the
ground $m = 0$ -component may be found from the following equation:

\begin{equation}
\frac{d}{dt}\rho _{00}=\Gamma _{00}\rho _{00}^{^{\prime }}\;.  \label{eq13}
\end{equation}

(ii) The initial conditions imply the absence of any coherence, i.e. $%
\rho_{+-}(t=-\infty) = 0$. It is clearly seen from Eq.~(\ref{10d}) that, if
a magnetic field is absent ($\omega_L = 0$) and $\widetilde{\Omega}_+%
\widetilde{\Omega}^*_-$ has a real value (i.e. the $x,y$-axes are the main
diagonals of the polarization ellipse), the imaginary part of the
off-diagonal element $\rho_{-+}$ remains zero.

(iii) The main feature of the case that occurs when $\omega_L = 0$ is that
the system (\ref{10a})-(\ref{10d}) has a unique simple stationary solution
which does not include the population of the excited state:

\begin{equation}
\begin{array}{l}
\rho _{00}^{^{\prime }}=0\;;\qquad \rho _{ii}=\frac{\widetilde{\Omega }_{-i}%
\widetilde{\Omega }_{-i}^{\ast }}{|\widetilde{\Omega }_{+}|^{2}+|\widetilde{%
\Omega }_{-}|^{2}}; \\
\rho _{ij}=-\frac{\widetilde{\Omega }_{i}\widetilde{\Omega }_{j}^{\ast }}{|%
\widetilde{\Omega }_{+}|^{2}+|\widetilde{\Omega }_{-}|^{2}}\qquad
(i,j=+,-)\;.%
\end{array}
\label{eq14}
\end{equation}


\subsection{\label{states} Dark states formed by spectrally broad light}


The density matrix in (\ref{eq14}) obviously corresponds to a pure quantum
state and allows one to determine an angle for mixing the $m=-1$ and $m=+1$
sublevels of the ground state, as in the case of coherent laser fields.
Since the excited state is not populated, the stationary solution (\ref{eq14}%
) describes a dark state that has been formed with spectrally broad light.
This result implies a beneficial cancellation of contributions from
incoherent frequency components of spectrally broad light and consequently
opens up new perspectives on the fruitful control of atomic states.

It is noteworthy that, under the absence of Zeeman separation ($\omega _{L}=0$), there
is only a single stationary state of the density matrix that allows $\rho
_{00}^{^{\prime }}=0$. To prove this statement, let us set the
left-hand-side of the system (\ref{10a})--(\ref{10d}) to be zero. Then Eqs.~(%
\ref{10b})--(\ref{10c}) for $\rho _{00}^{^{\prime }}=0$ yield

\begin{equation}
|\widetilde{\Omega }_{+}|^{2}\rho _{++}=- {\rm Re} \widetilde{\Omega }_{-}%
\widetilde{\Omega }_{+}^{\ast }\rho _{-+}=|\widetilde{\Omega }_{-}|^{2}\rho
_{--}  \label{eq15}
\end{equation}%
Since the system (\ref{10a})--(\ref{10d}) is a linear one, its solution may
be normalized by one parameter; we may choose $\rho _{++}$, for instance, in
the form $\rho _{++}=|\widetilde{\Omega }_{-}|^{2}/(|\widetilde{\Omega }%
_{+}|^{2}+|\widetilde{\Omega }_{-}|^{2})$. With such a choice, the relations
in (\ref{eq15}) acquire the form of (\ref{eq14}), which therefore turns out
to be unique. The normalization adopted in (\ref{eq14}) follows from the
requirement that $\rho _{++}+\rho _{--}=1$, which reflects the population
conservation. Significantly, the stationary case of Eqs.~(\ref{10a}),(\ref{10d}%
) (with zero left-hand-side) is satisfied automatically by the solution (\ref%
{eq14}) provided that $\omega _{L}=0$. The fact that the dark states formed
by spectrally broad light are unique fits well with our previous qualitative
considerations about $D$-states (see the discussion after Eq.~(\ref{eq9})).
In the Appendix we shall further motivate why the density
matrix in (\ref{eq14}) is a solution of the master equations (\ref{10a})--(%
\ref{10d}).

The density matrix in (\ref{eq14}) corresponds to the pure quantum state
with $\Psi$-function whose components have amplitudes $c_i$

\begin{equation}
\begin{array}{l}
c_+=\widetilde{\Omega}_-/\sqrt{|\widetilde{\Omega}_+|^2+|\widetilde{\Omega}%
_-|^2}; \\
c_-=-\widetilde{\Omega}_+/\sqrt{|\widetilde{\Omega}_+|^2+|\widetilde{\Omega}%
_-|^2}; \\
c^{^{\prime }}_0 = 0%
\end{array}
\label{eq16}
\end{equation}

In the case of a conventional $\Lambda $-scheme with two coherent lasers of
Rabi frequencies $\widetilde{\Omega }_{i}$ (which are complex in general),
the coherent dark state in (\ref{eq2}) has the same amplitude as determined
in (\ref{eq16}). Therefore, it is possible to analyze the situation with our
spectrally broad light or "white light" dark states ($WD$-states) using the
rich information known about standard coherent $D$-states. In particular, in
the next Section we will be able to express quantitatively to what extent
the state control process needs to be adiabatic in order to ensure the
survival of $WD$-states under non-stationary conditions.


\section{\label{whiteSTIRAP} spectrally broad light STIRAP}


There are a many interesting applications of the $WD$-state. First of all, since
the $WD$-state (\ref{eq14}) that arises in the scheme presented in Fig.~4
turns out to have the standard structure (\ref{eq16}) it would be beneficial
to consider the most recent developments in the applications of conventional
dark states \cite{Enio,Klaas} when exploring applications of the $WD$-state.


\subsection{\label{transfer} Population transfer}


We start by examining now the efficiency of the STIRAP-like population
transfer from the $m=-1$ component to the $m=+1$ component of the ground
state when the analogue of the two-photon resonance detuning $\delta
=2\omega _{L}$ is zero (see the clarifying remarks in the Appendix after
Eqs.~(\ref{A10a})--(\ref{A10d})). Initially, at ($t=-\infty $) the
population was found in the sublevel $m=-1$ in a pure state that corresponds
to

\begin{equation}
\begin{array}{l}
\rho_{--}=1 \; ; \qquad \rho_{ii}=0 \qquad (i \neq -) \; ; \\
\rho_{ij} = 0 \qquad (i \neq -j) \; .%
\end{array}
\label{eq17}
\end{equation}

We now consider how to drag most efficiently the population without loss from the $m=-1$
sublevel to the $m=+1$ sublevel by dynamically changing the properties of the light.
Based on previous experience with STIRAP
(Fig.~1), the mixing angle $\Theta $ (see Eq.~(\ref{eq2})) should be changed
from $0$ to $\pi /2$. Therefore, one has to modulate the polarization of the
light in such a manner that the corresponding effective Rabi frequencies
acquire pulse forms (\ref{11b}) in a necessary sequence: pulses are offset
in time, and in the pulse sequence the $\widetilde{\Omega }_{+}(t)$ pulse
should arrive first, i.e.,

\begin{equation}
\Delta \tau _{del}=\Delta \tau _{-}-\Delta \tau _{+}>0.  \label{eq18}
\end{equation}%
The pulses have duration $\tau $, and the temporal shift $\Delta \tau _{del}$
between them has to be positive.

The efficient population transfer (without loss) should be performed
adiabatically via $D$-states. Any deviation from adiabaticity results in
mixing between the state coupled to the excited state (the bright state) and
the $WD$-state that allows the population flow into unwanted $g$-level with $%
m=0$. The condition for implementing the adiabatic passage in the case of
coherent lasers was discussed above and is expressed by relation (\ref{eq5}%
). In the Appendix we derive a modification of criterion (\ref{eq5}) for
spectrally broad light excitation

\begin{equation}
\widetilde{\Omega }_{0}^{2}\tau \gg 1\;.  \label{eq19}
\end{equation}%
where the effective frequency $\widetilde{\Omega }_{0}$ is defined in Eq.~(%
\ref{11b}). This last requirement ensures the adiabaticity of STIRAP in the
case of system (\ref{10a})--(\ref{10d}).

\begin{figure}[tbp]
\begin{center}
\label{Fig5} \includegraphics[width=8cm]{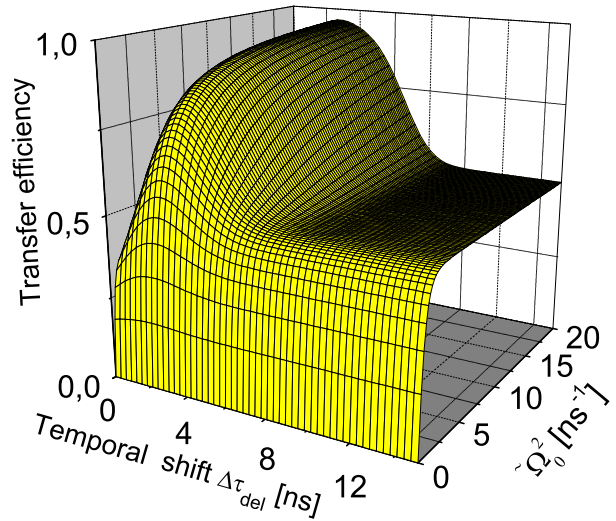}
\end{center}
\caption{Simulation of STIRAP with white light. The population transfer
efficiency was found from numerical solutions of system (\protect\ref{10a})-(%
\protect\ref{10d}) and is plotted versus the square of the effective Rabi
frequency $\widetilde{\Omega}^2_0$ [ns$^{-1}$] and the temporal shift $%
\Delta \protect\tau ^{}_{del}$ [ns] between the pulses. The radiative decay
constant $\Gamma _0$ =9 ns$^{-1}$ while the duration of pulses $\protect\tau$%
=2 ns.}
\end{figure}

In the Appendix we will examine as well the structure of the density matrix
equation (see Eqs.~(\ref{A10a})--(\ref{A10d})) for the case of two coherent
laser fields with a small two-photon detuning $\delta $ and a relatively large
one-photon detuning $\Delta $ (see Fig.~1). Such a one-photon detuning makes
it possible to adiabatically eliminate the optical coherence elements, which
reveals a close analogy between coherent (Eqs.~(\ref{A10a})--(\ref{A10d}))
and incoherent dynamics (Eqs.~(\ref{10a})--(\ref{10d})). With this analogy
in mind, intuitively one may expect to attain high efficiency with
spectrally broad light STIRAP. Fig.~5 illustrates the above conclusions by
showing the results of numerical simulations of population transfer. In this
example we fix the duration of the pulses to $\tau =2$ ns.
The transfer efficiency is measured as the population $\rho _{++}^{(f)}=\rho
_{++}(t=\infty )$ of the target state after the pulse sequence concludes.
Note that the surfaces of $\rho _{++}^{(f)}$, presented in Fig.~5 has
properties identical to the case of coherent lasers, as is shown in the
Appendix. The value $\rho _{++}^{(f)}$ is optimal, for instance, when $%
\Delta \tau _{del}\approx \tau $ \cite{Klaas}. The data exhibited in Fig.~5
illustrate as well criterion (\ref{eq19}) of transfer adiabaticity: if we
set a criterion for a successful population transfer at the level of $\rho
_{++}^{(f)}$ equal to $0.9$, the saturation starts in the region $\widetilde{%
\Omega }_{0}^{2}\tau >10$. If one wants to increase efficiency of population
transfer (increase value of $\rho _{++}^{(f)}$) further, one needs to
increase $\widetilde{\Omega }_{0}^{2}\tau $. Note that the simulation shows that a variation of the
decay constant $\Gamma _{0}$ by up to an order of magnitude does not
influence significantly the efficiency as a
function of the pulse area $\widetilde{\Omega }_{0}^{2}\tau $ and the delay $%
\Delta \tau _{del}$ between the pulses.


\subsection{\label{Zeeman} Influence of Zeeman splitting}


It is of particular interest for practical applications to examine what
happens when a weak magnetic field is present. The presence of Zeeman
splitting $\omega _{L}$ in system (\ref{10a})-(\ref{10d}) formally
corresponds to two-photon detuning with value $\delta =2\omega _{L}$ (see
Appendix). The $WD$-state (\ref{eq14}) fails to be a stationary solution of
system (\ref{10a})-(\ref{10d}), as the terms corresponding to $\omega _{L}$
mix the dark state with bright state, and some fraction of the $D$-state
population flows to unwanted states: $g$-state $m=0$ and the initial state $%
m=-1$. Figs.~6 and 7 give some insight into how the unwanted processes
change the desired transfer efficiency. We choose the situation with an
optimal delay $\Delta \tau _{del}=2$ ns in the pulse sequences and set the
decay constant to be $\Gamma _{0}=9$ ns$^{-1}$. Fig.~6 corresponds to the
case of state control with spectrally broad radiation, while Fig.~7 shows
the results of solving system (\ref{A10a})--(\ref{A10d}), that illustrates
conventional monochromatic STIRAP. In the latter case the effective
frequency $\bar{\Omega}_{0}$ is determined in Eq.~(\ref{A8}).

We point out for both Figs.~6,7 the somewhat curious behavior of $\rho
_{++}^{(f)}$ as a function of the light intensity in the region of large
two-photon detuning values. Initially when effective Rabi frequencies $%
\widetilde{\Omega }_{0}$ or $\bar{\Omega}_{0}$ are small, the curves $\rho
_{++}^{(f)}(\omega _{L}=\mathrm{con},\widetilde{\Omega }_{0})$ rise linearly
with increasing $\widetilde{\Omega }_{0}$ in the sequence of simple
observation: the population transfer occurs because the levels couple to
the light, where larger $\widetilde{\Omega }_{0}$ corresponds to stronger
coupling and, consequently, to larger transfer efficiency. This observation
is true as long as the interaction of an atom with photons is linear, i.e.,
light is unable to modify the bare states. When the light intensity starts
to exceed the saturation value, the bare states are transformed into dressed
states, each of which shares the population with the excited state (if $%
\delta \neq 0$) that clearly stimulates the unwanted population flow.
Moreover, the transformations are accomplished by energy shifts of the
dressed states that lead to some effects with similarities to laser induced
transparency. In the particular case of monochromatic STIRAP, because of the ac
Stark shift of the transition $m=-1\rightarrow m^{\prime }=0$, the initial
impulse $\bar{\Omega}_{+}(t)$ (see in Appendix Eq.~(\ref{A7})) results in
increasing detunings from the transition $m=+1\rightarrow m^{\prime }=0$.
Both aforementioned factors dramatically decrease $\rho _{++}^{(f)}$ in the
region $\widetilde{\Omega }_{0}^{2}\tau >18$ as follows from Figs~6,7.
The case of large $\widetilde{\Omega }_{0}$ values allows one to consider
the problem under a perturbation approach in which the parameter $\delta /%
\widetilde{\Omega }_{0}^{2}$ becomes small. One of the dressed states has a
structure close to that of a dark state (see Eq.~(\ref{eq14})). It shares only
a small fraction $\sim \delta /\widetilde{\Omega }_{0}^{2}$ of the excited
state and restores the population transfer efficiency in the region of very
large $\widetilde{\Omega }_{0}$.

\begin{figure}[tbp]
\begin{center}
\label{Fig6} \includegraphics[width=8cm]{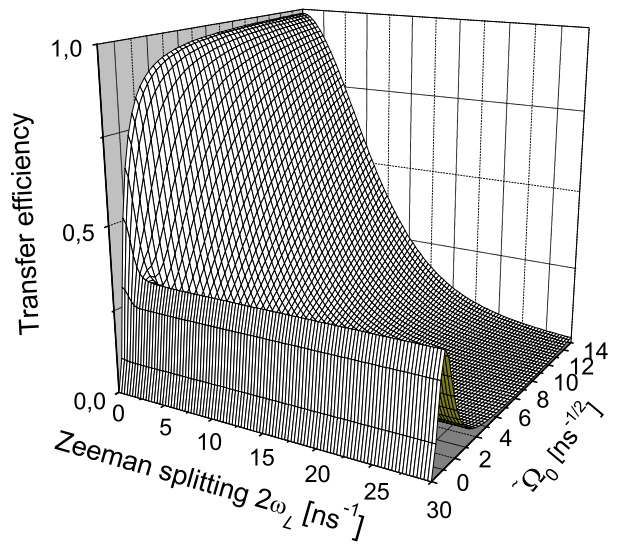}
\end{center}
\caption{Simulation of STIRAP with white light. Population transfer
efficiency versus the effective Rabi frequency $\widetilde{\Omega}^{}_0$ [$%
ns^{-1/2}$] and Zeeman splitting $\protect\omega_L$ [ns$^{-1}$]. The
temporal shift $\Delta \protect\tau ^{}_{del}$ between the pulses is chosen
to be 2 ns. The radiative decay constant $\Gamma _0$ =9 ns$^{-1}$ and the
duration of pulses $\protect\tau$=2 ns.}
\end{figure}

\begin{figure}[tbp]
\begin{center}
\label{Fig7} \includegraphics[width=8cm]{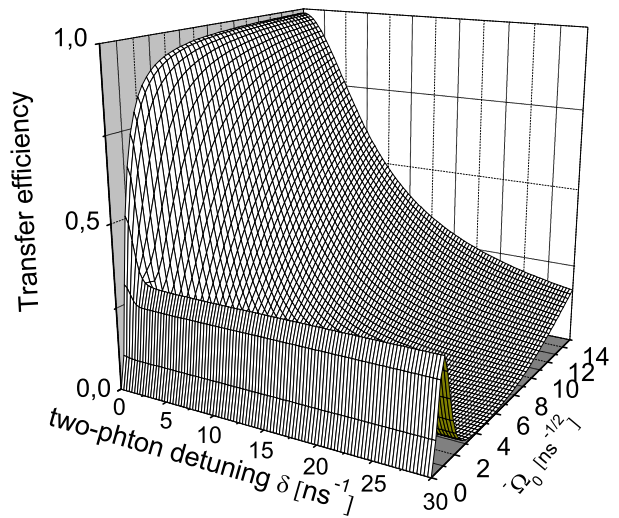}
\end{center}
\caption{Simulation of STIRAP with coherent lasers. Population transfer
efficiency versus the effective Rabi frequency $\bar {\Omega}^{}_0$ [$%
ns^{-1/2}$] and two-photon detuning $\protect\delta =2\protect\omega _L$ [ns$%
^{-1}$] (see in Appendix). The temporal shift $\Delta \protect\tau ^{}_{del}$
between the pulses is chosen to be 2 ns. The radiative decay constant $%
\Gamma _0$ =9 ns$^{-1}$ and the duration of pulses $\protect\tau$=2 ns.}
\end{figure}

\section{\label{concl} Conclusion}

In this paper we have demonstrated that if the states between which we carry
out population adiabatic transfer (STIRAP) are degenerate, this process can
be implemented with broad-band nonmonochromatic (\textquotedblleft
white\textquotedblright ) light. The efficiency of the population transfer
with broad-band radiation is similar to the efficiency of the STIRAP process
that can be achieved in the traditional way with monochromatic radiation and,
in the case of sufficiently slow, adiabatic manipulation of the states, can
approach 100\%.

The existence of dark states in the manifold of magnetic sublevels of an
atomic state created by the non-monocromatic radiation was noticed earlier
(see, for example, \cite{The89}) and was related to the well known phenomenon of
optical pumping of atoms in the manifold of magnetic sublevels \cite{Happer}%
. In $50$s, a long time before the invention of lasers, the phenomenon of optical
pumping was observed with a conventional spectrally broad light source by
Brossel, Kastler, and Winter \cite{Bro52} and by Hawkins and Dicke \cite%
{Haw53}. With this in mind, it should be easy to understand
the conclusions of this paper, that with
broad-band radiation it is possible not only to create a dark state in the
manifold of magnetic sublevels, but it is possible also to manipulate this state.

The demand that the states involved in the spectrally broad light STIRAP
process are degenerate is essential. It ensures that the phase fluctuations
of the radiation source are synchronous for the pump as well as the Stokes field and
cancels in the atom--light interaction process. Even a weak magnetic field
that splits magnetic sublevels on the order of the ground level-width
destroys the dark state created by the broad-band radiation.

It is obvious that the manipulation of coherent states created in the
manifold of the magnetic sublevels by the broad-band radiation is not
limited to the STIRAP process only, but can be extended to other coherent
processes, such as coherent control of atomic states with three
light fields in a tripod configuration \cite{Una2000} or manipulation of
many degenerate quantum states simultaneously \cite{Hei06}.

If one uses a broad-band radiation source with sufficient spectral density,
population transfer through the continuum \cite{Yat97} can be foreseen as
well. The advantage of white light in is that it does not significantly
perturb the continuum in contrast to the monochromatic lasers. The latter
strongly modify continuum states and result in sharp Fano profiles of dipole
matrix elements \cite{Fan61} that block the existence of stable dark states.



\begin{acknowledgments}
This work was supported by the EU FP6 TOK Project LAMOL (Contract
MTKD-CT-2004-014228), RFBR Grant No. 08-02-00136, by the Latvian Science
Council and the INTAS projects 06-1000024-9075 and 06-1000017-9001. We thank
Professor E. Arimondo, Professor K. Bergmann, Professor D.Budker, Dr. B.
Shore and Professor L. P. Yatsenko for the useful discussions.
\end{acknowledgments}

\appendix

\section{Density matrix equation with coherent lasers under adiabatic
conditions}

It is of interest to analyze the population transfer scheme presented in
Fig.~4 for the case of two coherent lasers. We assume a large enough
one-photon detuning value to ensure that the procedure of adiabatic
elimination (AE) of the optical coherence elements \cite{Stenholm} will be
valid. The same procedure is the main approach used in the broad-line
approximation to obtain system (\ref{10a})-(\ref{10d}) for the dynamics of
the density matrix driven with spectrally broad light. It is natural to
expect some similarity between the coherent and incoherent cases and to take
advantage of the well studied coherent case to predict important properties
of the manipulation of states with broad-band radiation.

We consider a three-level system, as depicted in Fig.~4, which is being
excited by two independent coherent laser fields with Rabi frequencies $%
\Omega _{\pm }(t)$. In an experiment it would mean that one applies lasers with
$\pm $ circular polarization to a two-level atom with angular momentum $l=0,1
$ for the upper and lower energy states, respectively. The ground state
Zeeman sublevel $m=0$ is not involved in the interaction with the light. It
collects the spontaneous population flow from the excited sublevel $%
m^{\prime }=0$ at the decay rate $\Gamma _{00}$. The lasers have identical
frequencies $\omega $, i.e., they are detuned from the $l^{\prime
}=0,m^{\prime }=0\leftrightarrow l=0,m=0$ Bohr transition frequency $\omega
_{0}$ at the same one-photon detuning $\Delta =\omega -\omega _{0}$. Because
of the possible presence of an external magnetic field, the Zeeman sublevels
$m=\pm 1$ may have Zeeman energy shifts with value $m\widetilde{\omega }_{L}$%
. Clearly, the two-photon detuning $\delta $ is then equal to the
corresponding Zeeman splitting $2\omega _{L}$ . If $\Delta $ happens to be
relatively large, for instance, if it effectively exceeds the inverse
duration $1/\tau $ of the laser pulses (we assume them to have Gaussian
shapes (\ref{eq9})), the adiabatic approximation becomes valid for the
density matrix elements $\rho _{m,m^{\prime }=0}^{{}}$ (the indices $m=\pm 1$
and $m^{\prime }=0$ belong to the ground and the excited states,
respectively) that make it possible to eliminate adiabatically $\rho
_{m,m^{\prime }=0}^{{}}$ \cite{Stenholm}. In addition, a large spontaneous
decay rate $\Gamma _{0}\tau \gg 1$ ensures that adiabatic
elimination is realized as well. Under the rotating-wave approximation \cite{Klaas}
adiabatic elimination allows us to set $d/dt\rho _{m,m^{\prime }=0}^{{}}=0$
and, thus, to reduce $\rho _{m,m^{\prime }=0}^{{}}$ to the form \cite%
{Stenholm, Blushs, Shore}

\begin{eqnarray}
\left( i\Delta _m+\frac 12\Gamma _0\right) \rho _{m,m^{\prime }=0}^{}(t)=
-\frac i2\Omega _m^{*}\rho _{m^{\prime }=0,m^{\prime }=0}^{}(t) \; + &&
\notag \\
\frac i2\Omega _{+}^{*}\rho _{m,+}^{}(t)+\frac i2\Omega _{-}^{*}\rho
_{m,-}^{}(t) \; . \quad &&  \label{A1}
\end{eqnarray}

\noindent The values $\hbar \Delta _m=\hbar (\Delta +m\omega _L^{})$ give
the energies of the $m$-bare states, provided that the energy of the $%
m^{\prime }=0$ bare state is chosen to be zero.

The coherent matrix elements $\rho _{m,m^{\prime }=0}^{}$ describe optical
oscillators in an atom with decay constant $\Gamma _0/2$. The oscillators
are excited by the lasers' radiation field, which have detunings $\Delta _m$. It is
well known \cite{Shore,Demtroder} that the excitation time $\tau ^{(ex)}$ is
determined by the relation $\tau ^{(ex)}\approx 1/\sqrt{\Delta _m^2+\Gamma
_0^2/4}$. If $\tau ^{(ex)}$ is substantially smaller than the
characteristic duration $\tau $ of the laser pulses, i.e.

\begin{equation}  \label{A2}
\tau \sqrt{\Delta _m^2+\Gamma _0^2/4}\gg 1 \; ,
\end{equation}

\noindent then the evolution of the amplitude $\rho _{m,m^{\prime }=0}^{{}}$
follows the excitation adiabatically, and Eq.~(\ref{A1}) comes to be
valid. This fact allows the general equation of motion for the density
matrix \cite{Shore, Blushs} to be reduced to the following system:
\begin{eqnarray}
\frac{d}{dt}\rho _{00}^{^{\prime }}=-\left( |\bar{\Omega}_{+}|^{2}+|\bar{%
\Omega}_{-}|^{2}+\Gamma _{0}\right) \rho _{00}^{^{\prime }}+|\bar{\Omega}%
_{+}|^{2}\rho _{++}\;+\qquad  &&  \notag  \label{A10d} \\
|\bar{\Omega}_{-}|^{2}\rho _{--}+2Re\bar{\Omega}_{-}\bar{\Omega}_{+}^{\ast
}\rho _{-+}\left( 1+\frac{2i\omega _{L}^{{}}}{\Gamma _{0}}\right) , \qquad
&&  \label{A10a} \\
\frac{d}{dt}\rho _{++}=-|\bar{\Omega}_{+}|^{2}\rho _{++}+\left( |\bar{\Omega}%
_{+}|^{2}+\Gamma _{0+}\right) \rho _{00}^{^{\prime }}\;-\qquad \qquad \quad
&&  \notag \\
Re\bar{\Omega}_{-}\bar{\Omega}_{+}^{\ast }\rho _{-+}\left( 1+\frac{2i\omega
_{L}^{{}}}{\Gamma _{0}}\right) , \qquad \qquad \qquad \qquad  &&
\label{A10b} \\
\frac{d}{dt}\rho _{--}=-|\bar{\Omega}_{-}|^{2}\rho _{--}+\left( |\bar{\Omega}%
_{-}|^{2}+\Gamma _{0-}\right) \rho _{00}^{^{\prime }}\;-\qquad \qquad \quad
&&  \notag \\
Re\bar{\Omega}_{-}\bar{\Omega}_{+}^{\ast }\rho _{-+}\left( 1+\frac{2i\omega
_{L}^{{}}}{\Gamma _{0}}\right) , \qquad \qquad \qquad \qquad  &&
\label{A10c} \\
\frac{d}{dr}\rho _{-+}=\left[ 2i\omega _{L}^{{}}-\left( |\bar{\Omega}%
_{+}|^{2}+|\bar{\Omega}_{-}|^{2}\right) \left( \frac{1}{2}+\frac{i\omega
_{L}^{{}}}{\Gamma _{0}}\right) \right] \rho _{-+}\;- &&  \notag \\
\frac{1}{2}\bar{\Omega}_{-}^{\ast }\bar{\Omega}_{+}\left( 1+\frac{2i\omega
_{L}^{{}}}{\Gamma _{0}}\right) \left( \rho _{++}+\rho _{--}-2\rho
_{00}^{^{\prime }}\right) , \quad  &&
\label{A10d}
\end{eqnarray}%
\noindent where we adopt the same notations for the density matrix elements
as in the case of broad-band radiation, Eqs.~(\ref{10a})--(\ref{10d}). In
particular $\rho _{00}^{^{\prime }}\equiv \rho _{m^{\prime }=0,m^{\prime
}=0}$. The equation for $\rho _{+-}$ is obtained from Eq.~(\ref{A10d}) by
complex conjugation. Two new effective "frequencies" $\bar{\Omega}_{\pm }$
are introduced
\begin{eqnarray}
\bar{\Omega}_{\pm }(t) &=&\frac{\sqrt{\Gamma _{0}}}{\sqrt{\Gamma
_{0}^{2}+4\Delta _{\pm }^{2}}}\Omega _{\pm }(t)\;;  \notag  \label{A7} \\
\Omega _{\pm }(t) &=&\Omega _{0}\exp (-(t-\Delta _{\pm })^{2}/2\tau ^{2})
\end{eqnarray}%
which have dimension [s$^{-1/2}$] and Gaussian pulse shapes of the type (\ref%
{11b}), which have arrived, as before, in an appropriate sequence.

Systems (\ref{A10a})--(\ref{A10d}) and (\ref{10a})--(\ref{10d}) provide
useful insight into how to apply knowledge of state control with two
coherent fields to the case of state control with spectrally broad
radiation. For instance, the reason for the existence of dark sates (\ref%
{eq14}) becomes clear, namely, when Zeeman splitting is absent (i.e. $\omega
_{L}^{{}}=0$ ), systems (\ref{A10a})--(\ref{A10d}) and (\ref{10a})--(\ref%
{10d}) become identical (provided we identify the effective Rabi frequencies
$\widetilde{\Omega }_{\pm }$ and $\bar{\Omega}_{\pm }$ ) and the coherent
dark state (\ref{eq2}) generates a spectrally broad light dark state in the
form of (\ref{eq14}). Note that this situation corresponds to two-photon
resonance when $\Delta _{+}=\Delta _{-}=\Delta _{{}}$. It is clear as well
that criterion (\ref{eq5}) for STIRAP adiabaticity is reduced to the
requirement $\Omega _{0}^{2}\tau ^{2}\gg 1$ for the pulses $\Omega _{\pm }(t)
$ (\ref{A7}). It is more instructive to express this requirement in terms of
the effective value $\bar{\Omega}_{0}$ for Rabi frequency:
\begin{eqnarray}
\bar{\Omega}_{0} = \Omega _{0}^{{}}\sqrt{\Gamma _{0}}/\sqrt{\Gamma
_{0}^{2}+4\Delta _{{}}^{2}}\;;  \notag  \label{A8} \\
\bar{\Omega}_{0}^{2}\tau \left( \tau \sqrt{\Gamma _{0}^{2}+4\Delta _{{}}^{2}}%
\right) \sqrt{1+4\Delta _{{}}^{2}/\Gamma _{0}^{2}} &\gg &1\;.
\end{eqnarray}%
\noindent Since system (\ref{10a})-(\ref{10d}) was obtained under
assumption (\ref{A2}), relation (\ref{eq19}) (with the clear substitution $%
\widetilde{\Omega }_{0}\rightarrow \bar{\Omega}_{0}$) turns out to be
sufficient to satisfy inequality (\ref{A8}). In other words, if we identify
the effective Rabi frequency $\widetilde{\Omega }_{0}$ with $\bar{\Omega}_{0}
$, we obtain a new criterion (\ref{eq19}) for efficient population transfer
with spectrally broad light under the realization of two-photon resonance.

\end{document}